%% file: main_inc.tex
\pdfoutput=1
\documentclass[12pt,
				a4paper,
				twoside,
				appendixprefix,
				BCOR8mm]{scrbook}

\usepackage{color}
\usepackage{pdfpages}
\usepackage{graphicx}

\usepackage{tabularx}
\usepackage{booktabs} 

\usepackage[numbers]{natbib}

\usepackage[ngerman,english]{babel}
\usepackage{csquotes}

\usepackage{user}

\usepackage[hidelinks]{hyperref}

\begin{document}

\pagestyle{plain}

\include{metadata}
\include{title}

\catcode`_=\active
\newcommand_[1]{\ensuremath{\sb{\mathrm{#1}}}}
\catcode`^=\active
\newcommand^[1]{\ensuremath{\sp{\mathrm{#1}}}}

\chapter{Introduction}
\label{chap:Intro}

Synchrotron light sources are of vital importance to modern spectroscopy and surface science.
Due to their unique properties combination of uniformly high brightness and brilliance over a wide frequency range, synchrotron light sources find broad application in solid-state surface science, chemistry, biology and life science\cite{Bilderback2005}.

In a linear accelerator based light source, the properties of the electron beam that emits synchrotron radiation are largely defined at the first stages of beam production and acceleration.
Especially, for prospected future light sources, such as X-Ray Free Electron Lasers (FELs) and Energy Recovery Linacs (ERLs), initial beam parameters are key factors of performance because no damping as in storage rings occurs.
In an effort to explore research and development areas required for future ERL design and operation, the demonstration facility \berlinpro~will be built at Helmholtz Zentrum Berlin.
It will  demonstrate feasibility of the ERL concept in a parameter scale envisaged for X-ray facilities.

The superconducting radio frequency (SRF) photo electron gun is one promising concept to deliver electron beams of the desired quality of ultra-low emittance below 1\,mm\,mrad and high average current in the order of 100\,mA.
Gun 0.2 is a demonstration project to explore the generation of a low current beam from a photoinjector where cathode, cavity and solenoid are all superconducting.
Cavity operation at high gradients was demonstrated, albeit at low beam loading and a low duty cycle.

In order to investigate and understand the influence of gun design, cathode preparation and operational parameters on the beam quality it is important to have reliable and accurate beam diagnostics available.
The focus of this work was thus to employ the available diagnostic beam line of the current gun demonstrator to characterize the transverse phase space of the beam and measure the emittance at various beam settings and operation conditions.
Slit mask and solenoid scanning techniques are applied.

\chapter{Theoretical Background}
\label{chap:Theory}

The figure of merit for the performance of a light source is the photon beam's brightness $B$. It is defined as the photon flux $F$ of a given bandwidth divided by the electron phase space volume that emits the photon flux:

\begin{equation}
B = \frac{F}{4\pi \sigma_x \sigma_y \sigma_x^\prime \sigma_y^\prime}.
\end{equation}
Which for a round beam (with no correlation) equals
\begin{equation}
B = \frac{F}{4\pi \varepsilon_x \varepsilon_y},
\end{equation}
where $\sigma$, $\sigma^\prime$ and $\varepsilon_{x,y}$ are the transverse width, divergence and emittance, respectively, of the electron beam in x or y direction. In a linac, the beam emittance depends heavily on the emittance of the gun. Thus, it is an important quantity that determines the brightness achievable by a linac driven light source.

\section{Transversal Phase Space and Emittance}

Transverse particle motion in a co-moving coordinate system can be described by the Courant-Snyder invariant

\begin{equation}
\label{eq:courant-snyder} \gamma u^2 + 2\alpha u u^{\prime} +\beta u^{\prime \, 2} = \varepsilon
\end{equation}
where $\alpha = -\frac{1}{2} \beta^{\prime}$, $\gamma = \frac{1+\alpha^2}{\beta}$. The parameters $\alpha, \beta$ and $\gamma$ are called Twiss Parameters. $u$ and $u^\prime$ are position and divergence in one transverse direction. Rigour derivations can be found in the text books by Wille, Wiedemann or Reiser \cite{WilleEngl, Wiedemann2007, Reiser2008}.

As can be seen from (\ref{eq:courant-snyder}), a particle's betatron oscillations describe an ellipse in transverse phase space while it moves through the focussing magnetic structure of an accelerator. The geometric emittance of a single particle is defined as the area of this ellipse up to a factor of $\pi$.

A beam, however, may consist of several $10^9$ electrons per bunch. It is therefore practical to reference emittance not to single particles but rather to a statistical ensemble. One common definition of the geometric beam emittance is thus the phase space area populated by all particles.

\pagebreak

Since the beam quality often degrades by filamentation while the populated phase space area remains constant due to Liouville's theorem it is feasible to define a geometric rms-emittance through the second moments of the beam's particle density distribution as

\begin{equation}
\label{eq:eps-definition}\varepsilon_{rms,\, g} = \sqrt{<u^2>\,<u^{\prime \, 2}> - <u \, u^{\prime}>}.
\end{equation}

In case of an elliptical beam the two definitions are equivalent if the same percentage of particles is enclosed in the ellipse. They are also referred to as the \textit{projected} transverse emittance, since correlations with the longitudinal coordinate are not considered.

The normalized emittance 
\begin{equation}
\label{eq:normeps-definition}\varepsilon_{n} = \beta \gamma \varepsilon_g
\end{equation}
remains constant during acceleration and allows comparison between beams of different energies.

\chapter{Measurement of the Emittance}
\label{chap:measuring-emittance}

Two techniques to measure the emittance are discussed: both yield the aforementioned rms emittance, one by summing up second moments of the particle distribution, the other by fitting parameters that depend on the rms distribution. Statistical information about the beam is obtained by peak finding algorithms that are applied to the measured intensity distribution.

\section{Slit Based Technique}
\label{chap:slit}

\begin{figure}[ht]
\centering
\includegraphics[width=.75\textwidth]{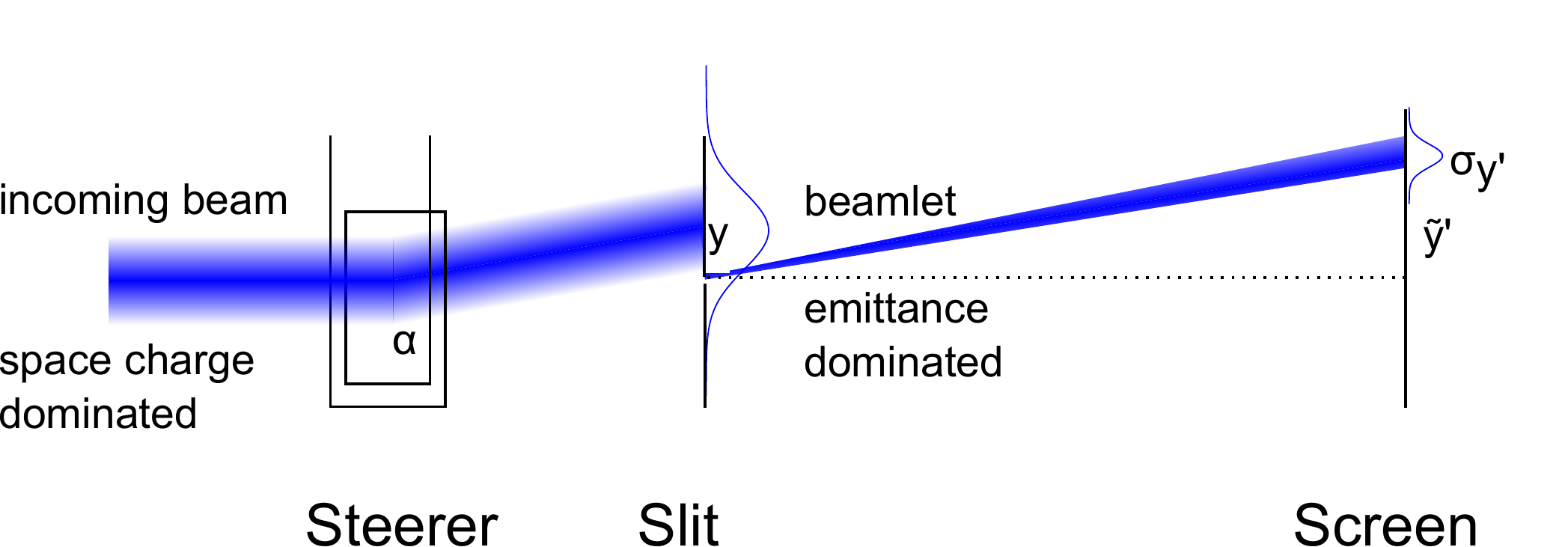}
\caption{Schema of the measurement setup for a slit mask measurement.}
\label{fig:scheme-slit-mask}
\end{figure}

The slit mask measurement allows to reconstruct the projected phase space distribution in one transverse direction at the location of the slit, as illustrated in figure \ref{fig:scheme-slit-mask}. The slit serves two purposes. For one, the slit selects a narrow band of the particle distribution, which allows to scan the beam's diameter and correlate transverse position and divergence. The width and mean value of the divergence of the beamlet are calculated from the transverse distance travelled in a drift path. Furthermore, if the beam dynamics upstream of the slit is space charge dominated, the aperture reduces the bunch charge so one may assume emittance dominated dynamics in the downstream beamlet. Space charge effects on the beamlet are neglected in this work, but do have an influence on the measured emittance at higher bunch charges \cite{Bazarov2008}.

\begin{figure}[ht]
\centering
\includegraphics[width=.75\textwidth]{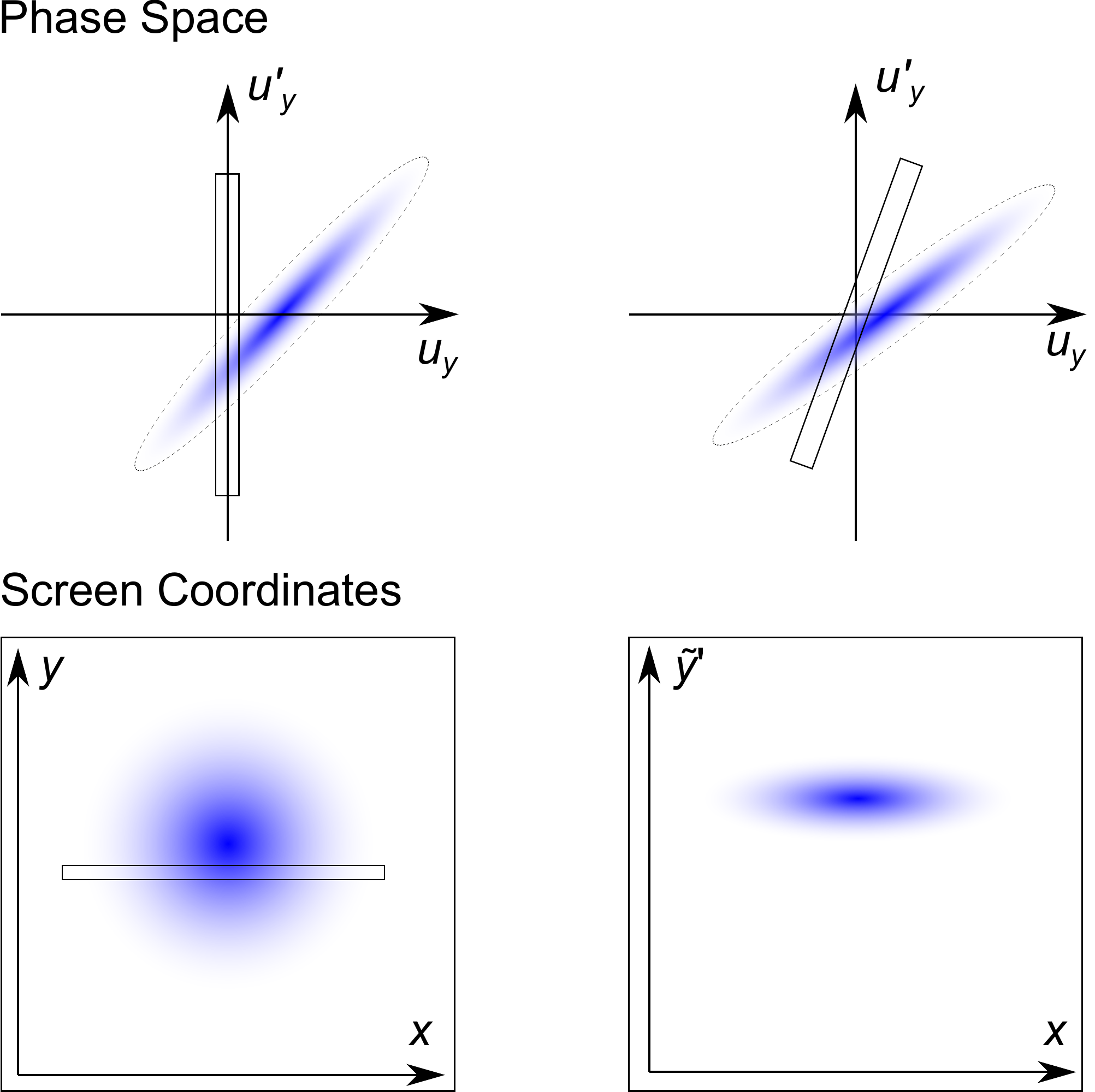}
\caption{Illustration of real and phase space distributions during a slit mask measurement.}
\label{fig:real-and-phase-space}
\end{figure}

The emittance can be evaluated using equation \ref{eq:eps-definition} and the second moments as noted hereafter.
$y$ and $y'$ denote spatial positions of particles on a screen, $u_y$ and $u_y^\prime$ denote phase space coordinates for vertical position and divergence as indicated in figure \ref{fig:real-and-phase-space}:

\begin{equation}
\langle u_y^2 \rangle_{Beam} = \frac{\sum\limits_i^{Particles} y_{i}^2}{\sum\limits_i^{Particles} 1} = \frac{\sum\limits_j^{Beamlets} I_j \langle y^2 \rangle_{j}}{\sum\limits_j^{Beamlets} I_j}
\end{equation}

\begin{equation}
\langle u_y^{\prime 2} \rangle_{Beam} = \frac{\sum\limits_i^{Particles} y_i^{\prime 2}}{\sum\limits_i^{Particles} 1} =
 \frac{\sum\limits_j^{Beamlets} I_j  \langle y^{\prime 2} \rangle_j}{\sum\limits_j^{Beamlets} I_j} =
 \frac{\sum\limits_j^{Beamlets} I_j ( \mu_j^2 + \sigma_j^2 )}{\sum\limits_j^{Beamlets} I_j} 
\end{equation}

\begin{equation}
\langle u_y u_y^{\prime} \rangle_{Beam} = \frac{\sum\limits_j^{Beamlets} I_j \langle y\rangle_j \langle y^{\prime}\rangle_j}{\sum\limits_j^{Beamlets} I_j}
\end{equation}

here, $I_j$ is the measured intensity of the $j$-th beamlet on the screen, $\mu_j$ and $\sigma_j$ are center and standard deviation of an assumed normal distribution of the individual divergences $y^{\prime}_i$ in one beamlet.\\
Note, that in case of steering $y$ and $\mu_j$ need to be corrected for the collective divergence due to beam steering:

\begin{equation}
y^{\prime} = \frac{\tilde{y}^{\prime} - \alpha L}{l_{beamlet}}
\end{equation}

\begin{equation}
\mu_j = \langle y^{\prime} \rangle = \frac{\langle \tilde{y}^{\prime} \rangle - \alpha L}{l_{beamlet}}
\end{equation}

where $\alpha$ is the steering angle, $l_{beamlet}$ is the drift length of the beamlet between slit mask and screen and $L$ is the length between steerer and screen. $\tilde{y}^{\prime}$ denotes the measured position on a screen, that is shifted by the steerer angle.

\clearpage
\pagebreak
\section{Solenoid Scanning}

Using the solenoid to scan the beam waist through a screen yields information about the initial beam size, divergence and their correlation.
Photoinjectors are equipped with an emittance compensating solenoid that is used for these measurements.
In a linear beam optics model, the beam radius at the screen depends quadratically on the solenoid's focal length \cite{myVolker2012},  as illustrated in figure \ref{fig:scheme-solenoid-scan}b:

\begin{figure}[ht]
\centering
\includegraphics[width=.75\textwidth]{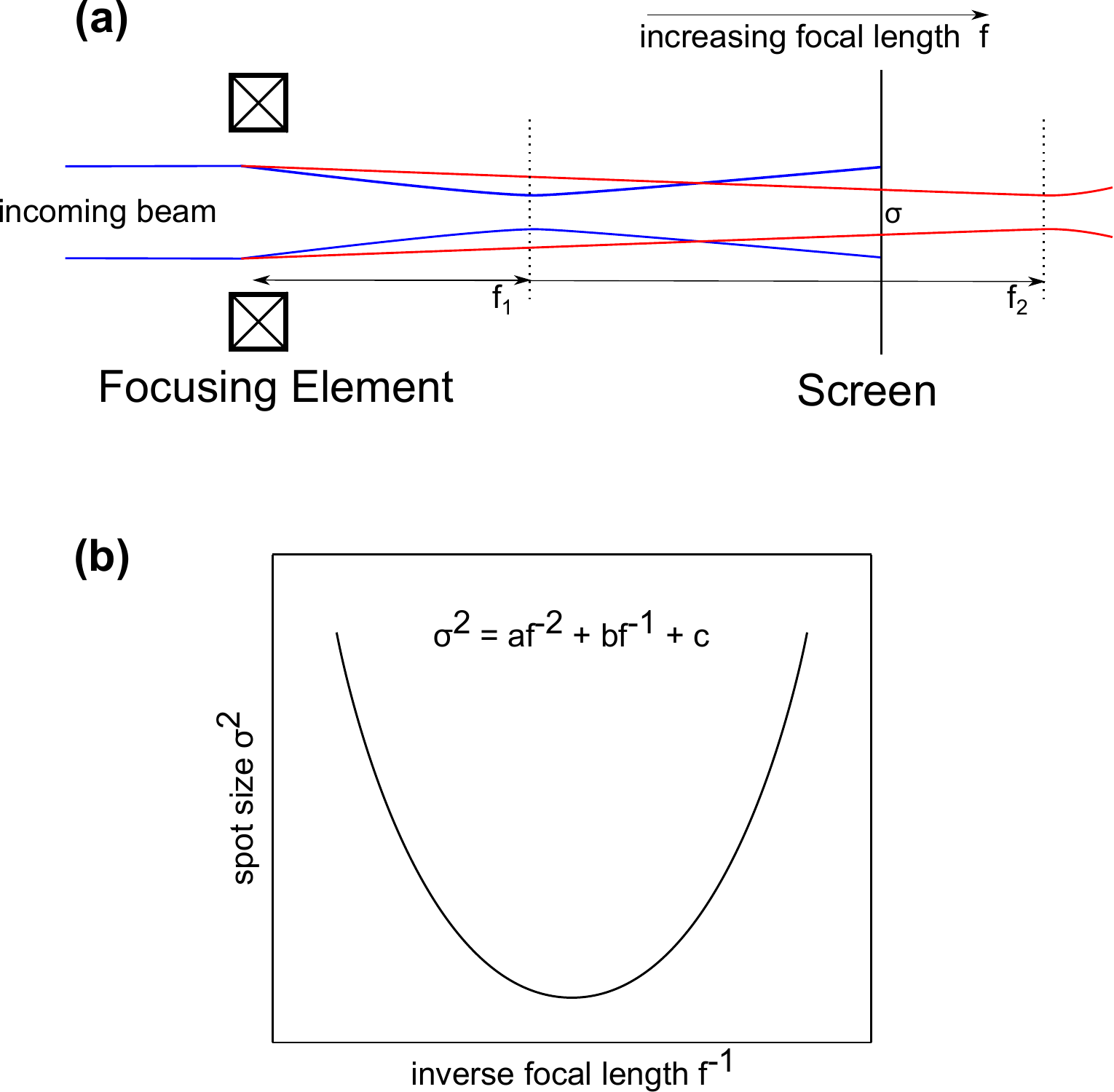}
\caption{Schema of the measurement setup for a solenoid scan (a) and illustration of the quadratic fit to obtain parameters for the emittance calculation (b).}
\label{fig:scheme-solenoid-scan}
\end{figure}

\begin{equation}
\langle r_{f,L}^2 \rangle = \langle ( r_i + r_f^\prime \cdot L)^2 \rangle
\end{equation}

\begin{equation}
r_f^\prime = r_i^\prime - \frac{r_i}{f}
\end{equation}

\begin{equation}
\langle r_{f,L}^2 \rangle = \frac{1}{f^2} \underbrace{(\langle r_i^2 \rangle L^2)}_a + \frac{1}{f} \underbrace{(-2L(L\langle r_i r_i^\prime \rangle + \langle r_i^2 \rangle))}_b + \underbrace{\langle r_i^{\prime 2} \rangle L^2 + 2L\langle r_i r_i^\prime \rangle + \langle r_i^2 \rangle}_c
\end{equation}

\begin{equation}
\varepsilon_{rms} = \frac{\sqrt{ac-\frac{b^2}{4}}}{L^2}
\end{equation}
where L is the distance between solenoid and screen, $r$ and $r'$ are the radial position and divergence of a particle at initial (before solenoid) and final (screen) positions, indicated by an i or f subscript, respectively.
The parameters $a, b$ and $c$ can be obtained from a quadratic fit of the beam size, plotted against the inverse focal lengths of the solenoid.

\section{Numerical Tracing of Particle Trajectories}

Numerical tracing of particle trajectories with ASTRA \cite{Astra} has been used to validate the analysis scripts used for the measured slit mask data. Further investigations regarding the influence of slit aperture and distances on measurement accuracy of the slit mask measurements are envisaged.
Particle tracking was also employed to obtain an estimate on the momentum spread of the beam, which was not possible to measure with the current setup but is required to estimate the chromatic aberration of the solenoid.

\chapter{Experimental Setup}
\label{chap:Setup}

An SRF injector is driven by a laser source that extracts photoelectrons from a cathode. The cathode is located in the back wall of an RF cavity where the electron bunches are accelerated by a strong electric field gradient, timed to match the correct phase of the RF wave.
The beam is fed into a diagnostics beam line, where fundamental beam parameters can be measured.

\begin{figure}[ht]
\centering
\includegraphics[width=.75\textwidth]{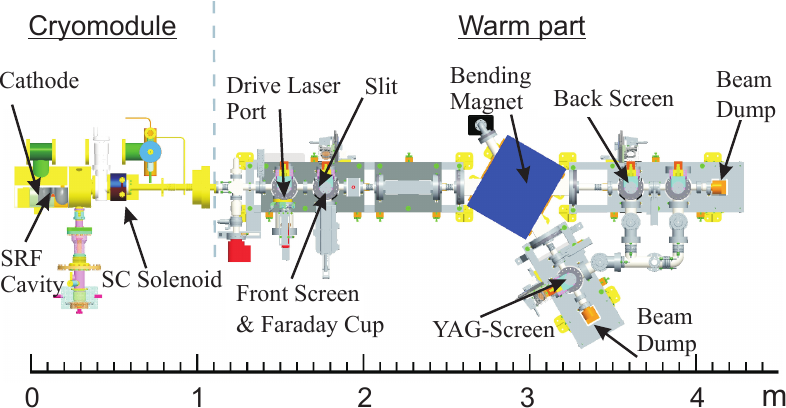}
\caption{Overview of the HoBiCaT cryomodule and the diagnostic beamline.}
\end{figure}

\section{Electron Gun}

The electron gun is the second prototype ("Gun 0.2") in a step-by-step approach towards the \berlinpro~injector.
It consists of a 1.6 cell superconducting niobium cavity that is located in the HoBiCaT cryostat, at an operational temperature of 1.8\,K, and a cathode plug inserted into the back wall of the cavity.
The cavity was fed with microwave power from an inductive output tube (IOT) or a solid state amplifier at 1.3\,GHz and was operated at field gradients up to 28\,MV/m, limited by field emission followed by slow quench. Stable operation was possible up to 27\,MV/m.
The cathode plug was made of niobium covered with a thin lead film. Lead has an approximately one order of magnitude higher quantum efficiency but is also superconducting below 7.2\,K.
Table \ref{tab:gun-parameters} lists typical operational parameters of the gun.

\begin{table}[!ht]
\centering
\setlength{\arrayrulewidth}{0.25pt}
\begin{tabular}{c c c}
\hline 

\textbf{Parameter} & \textbf{Value} & \textbf{Unit}    \\ \midrule

Average Current    & $<$ 1          & nA               \\ \midrule
Bunch Charge       & 0.187          & pC               \\ \midrule
QE                 & $10^{-5}$      &                  \\ \midrule
Beam Energy        & 1-2.5          & MeV              \\ \midrule
Laser Power        & $<$ 0.5        & mW               \\ \midrule
Laser wavelength   & 258            & nm               \\ \midrule
Pulse Length       & 2.5 \ldots 3   & ps fwhm, gaussian\\ \midrule
Rep rate           &  8             & kHz              \\ \midrule
$\mathrm{E_{max}}$ & \parbox{5cm}{\centering 10 - 12.5 w solid amp\\ 22 with IOT\\ 27 peak field} & MV/m  \\ 
\hline 
\end{tabular}
\caption{Typical operational parameters of Gun 0.2 with a lead coated cathode. The pulse length refers to the emission time of the bunch. }
\label{tab:gun-parameters}
\end{table} 

\section{Diagnostic Beam Line}

Several screens were employed to image beam profiles and to aid beam positioning. 
The solenoid can be used to focus the beam on the front or back screens in the straight beam line,
where a solenoid scan yields emittance information as described in section \ref{chap:measuring-emittance}.
A superconducting solenoid with an effective magnetic length of 41.5\,mm and a field amplitude of $44.1\,\frac{\mathrm{mT}}{\mathrm{A}}$ \cite{myVolker2012} was located inside the cryomodule.
The distance between the cathode and the center of the solenoid was 439\,mm.
Three beam stops made of copper served as faraday cups for current measurements. They were located at the first screen station and at the two ends of the beamline.

The slit at the first screen station was used to select a narrow beamlet which was imaged on the back screen.
The slit mask was made of 1.5\,mm thick tungsten with an $100\,\mu$m aperture and located directly in front of the first screen and faraday cup. A thickness of 1.5\,mm gives a good compromise between efficient suppression of the background and sufficient transmission.

In the dispersive section, after the beam has passed the dipole, it was possible to evaluate the momentum of the particles. Because the dipole could not be driven into saturation it was calibrated by cycling between +8\,A and -8\,A \cite{myMatveenko}. The beam momentum can be calculated as

\begin{equation}
pc = 0.88 \cdot \left( \frac{I_{Dipole}}{[A]} + 0.12 \right) [MeV].
\end{equation}

\section{Procedures}

Three individual measurements were conducted to characterize the phase space of one beam setting using the slit mask. First, the beam was imaged on the front screen without slit to calibrate the steerer angle and vertical offset. In a second sweep the front screen was used again, but with the slit to measure the beamlet intensity. Finally, the beamlets were allowed to drift towards the back screen in order to measure the divergence of each beamlet.
All measured values were taken from ten polls per beamlet on the back screen and five polls per beamlet on the front screen. Mean and standard deviation were recorded.
The phase space distributions were corrected for the steerer angle and summed up to obtain second moments and the emittance as described in \cite{Anderson2002}.

In order to conduct a solenoid scan, the current in the solenoid was varied over a specified range \cite{myVolker2012}.
Screen images were saved for a single poll per solenoid setting,
while beam positions and diameters were averaged over six polls. 
Beam parameters and the emittance are obtained from a quadratic fit of the beam width against the inverse focal length of the SC solenoid.

\chapter{Results}
\label{chap:results}

All measurements were conducted in November 2012 during the last run of the "Gun 0.2" setup. At that time, the cavity was operated with full RF power from the IOT only for test measurements and energy calibration. Due to a failure in the power supply of the IOT only up to 12.5\,MV/m were available for the subsequent beam measurements using the solid state amplifier.

\section{Slit Mask Measurements}

\begin{figure}[ht]
\centering
\includegraphics[width=.75\textwidth]{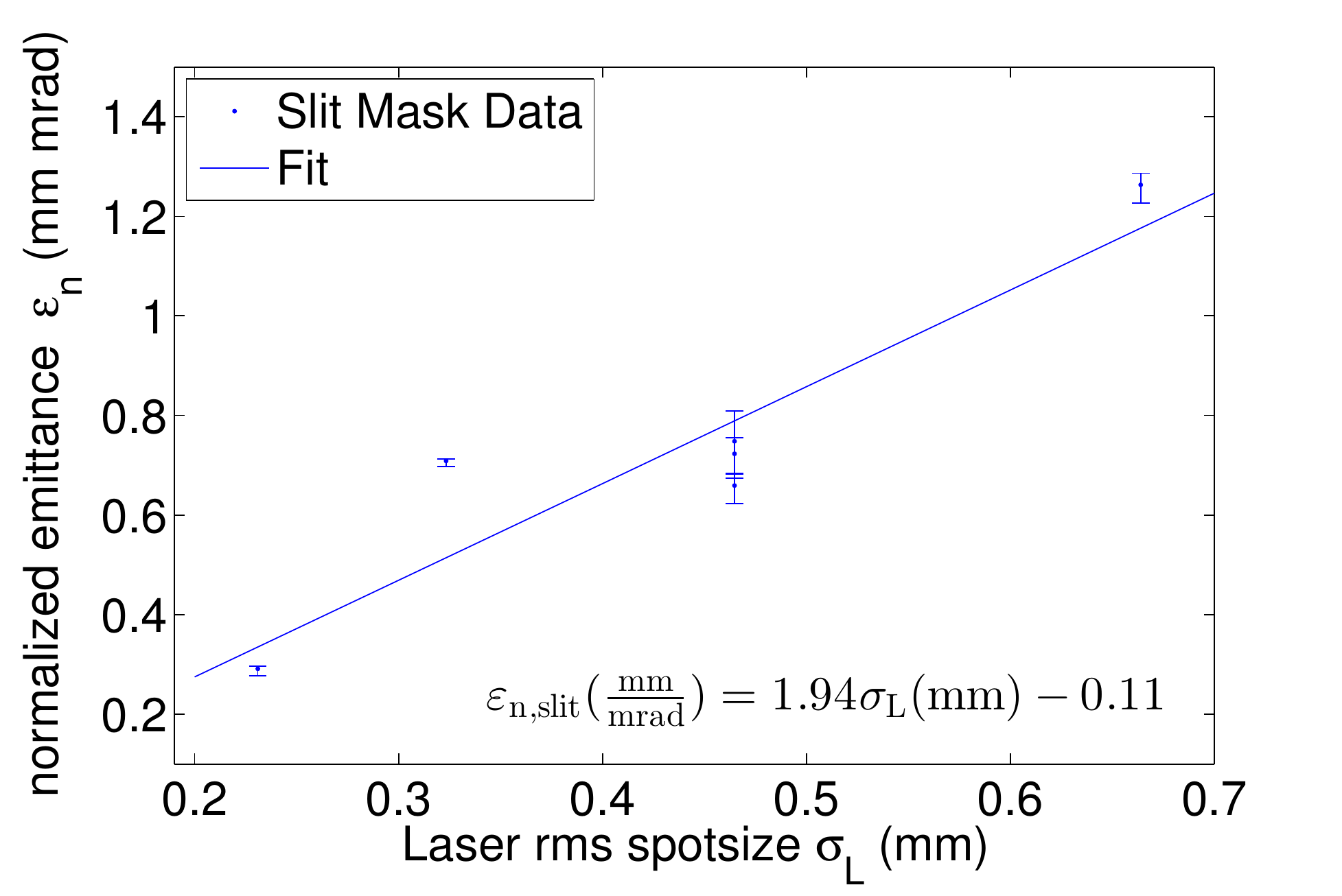}
\caption{Results of emittance measurements using the slit mask technique. Bunch charges are not equal.}
\label{fig:slit-mask-results}
\end{figure}

The vertical phase space was characterized at different laser spot sizes with a constant cavity gradient of 10\,MV/m at an emission phase of approximately 15\,deg using the slit mask as described in section \ref{chap:slit}. An aperture in the laser beam line was used to manipulate the spot size, however, other laser settings remained constant so the bunch charges were not equalized among the measurements (but all below the space charge regime). At a spot size of 0.46\,mm\,rms three solenoid settings were studied. The resulting emittance values are displayed in figure \ref{fig:slit-mask-results}.

\begin{figure}[ht]
\centering
\mbox{\includegraphics[width=.5\textwidth]{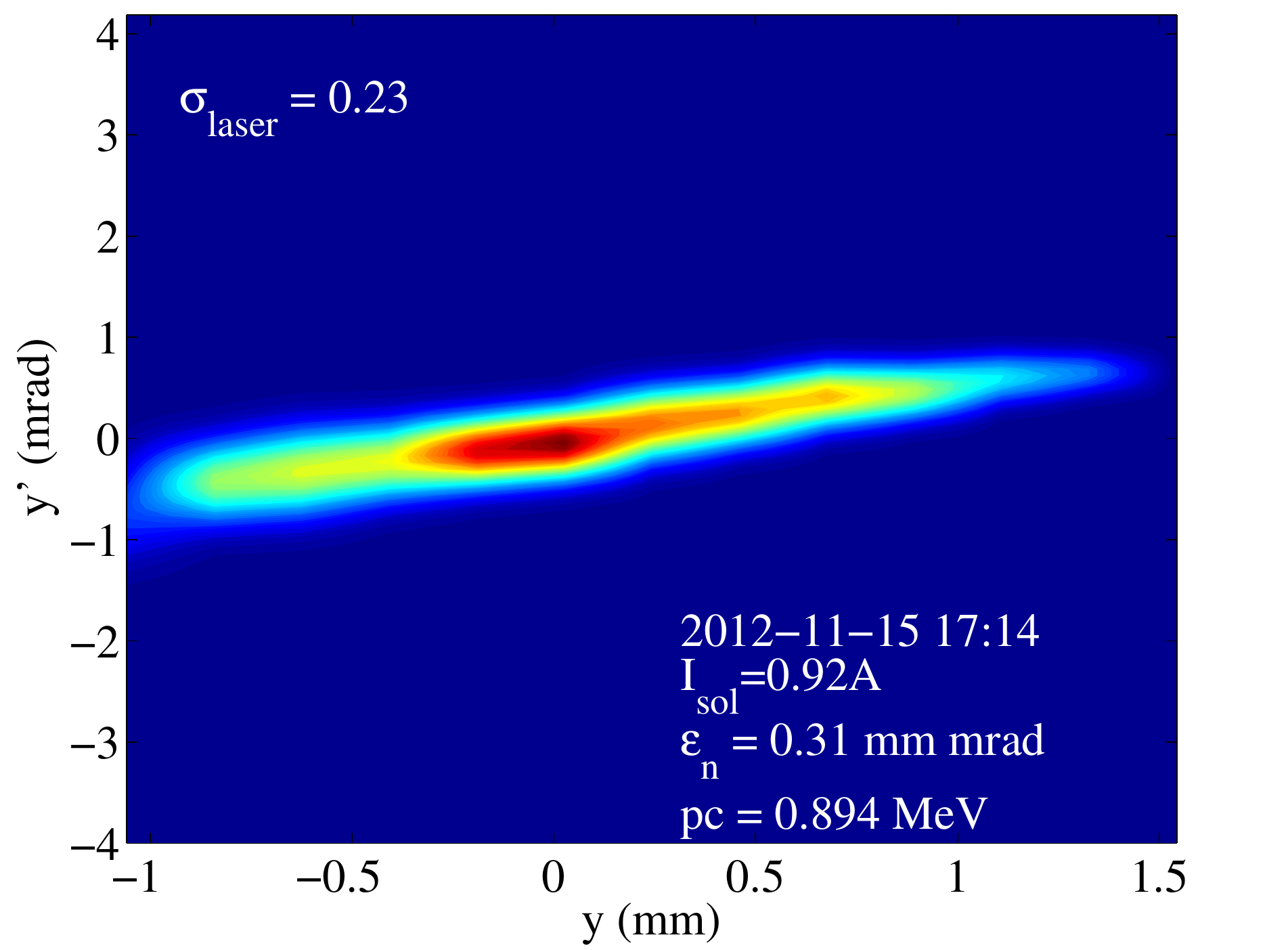}
\includegraphics[width=.5\textwidth]{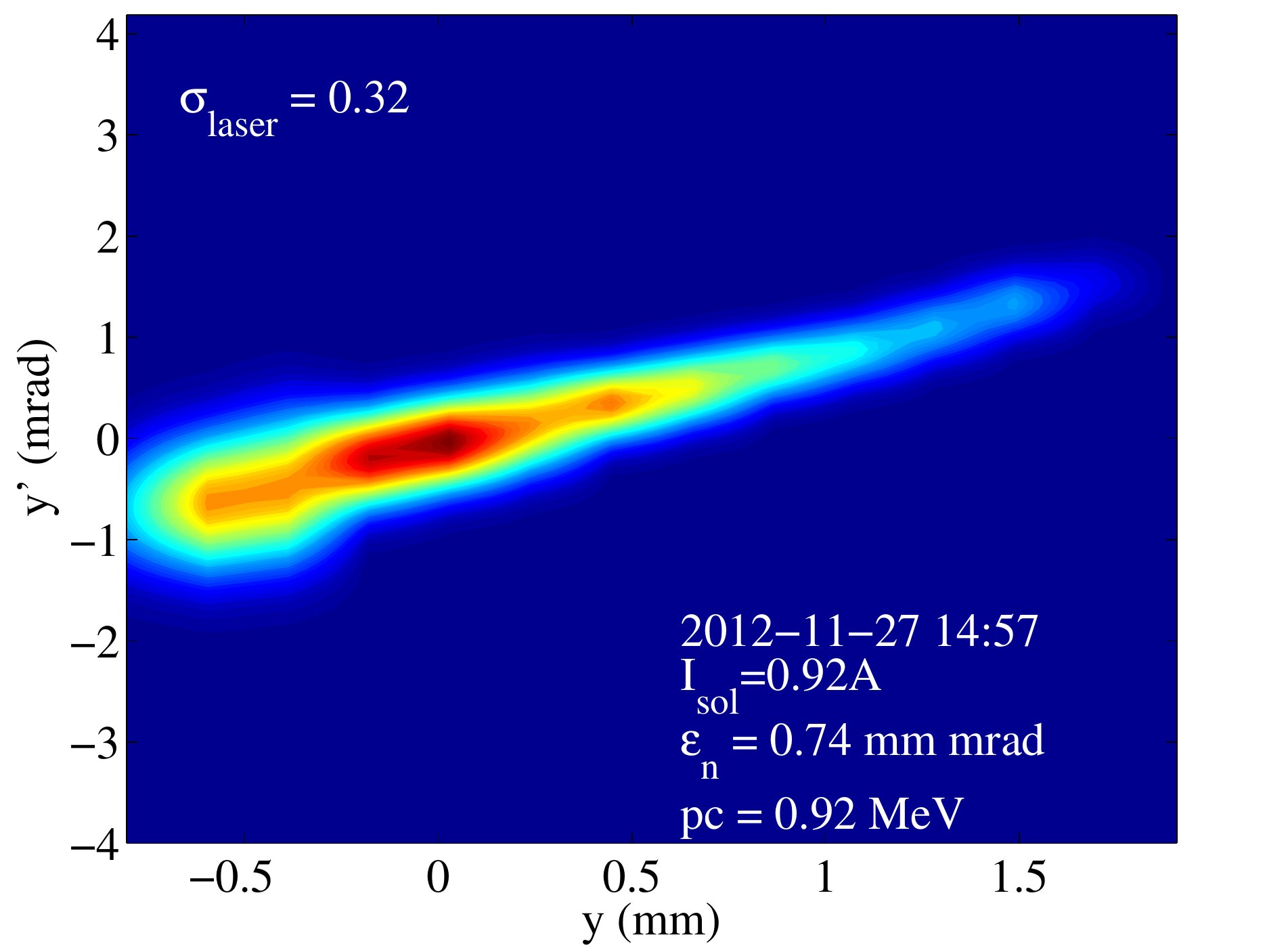}}
\mbox{\includegraphics[width=.5\textwidth]{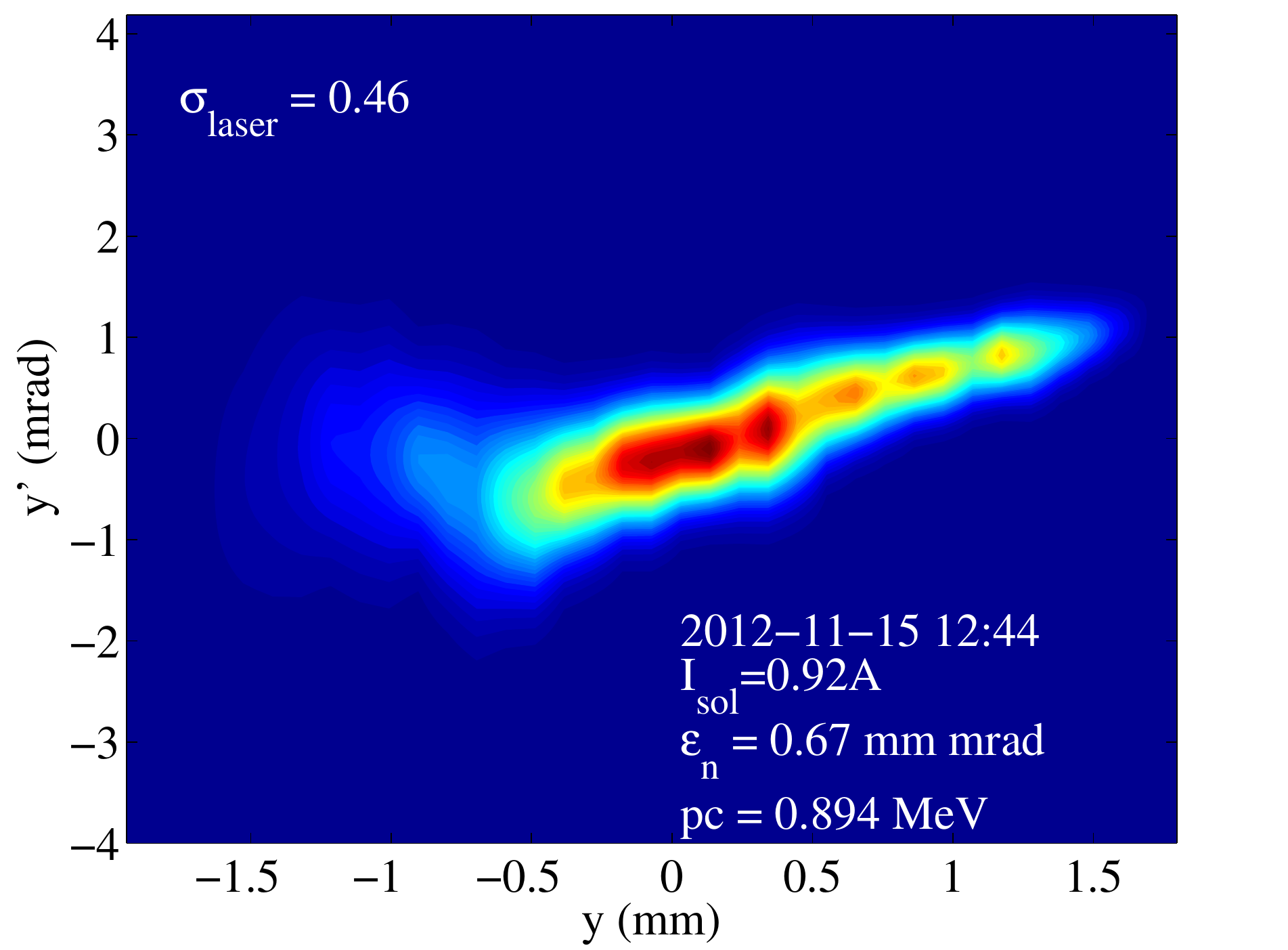}
\includegraphics[width=.5\textwidth]{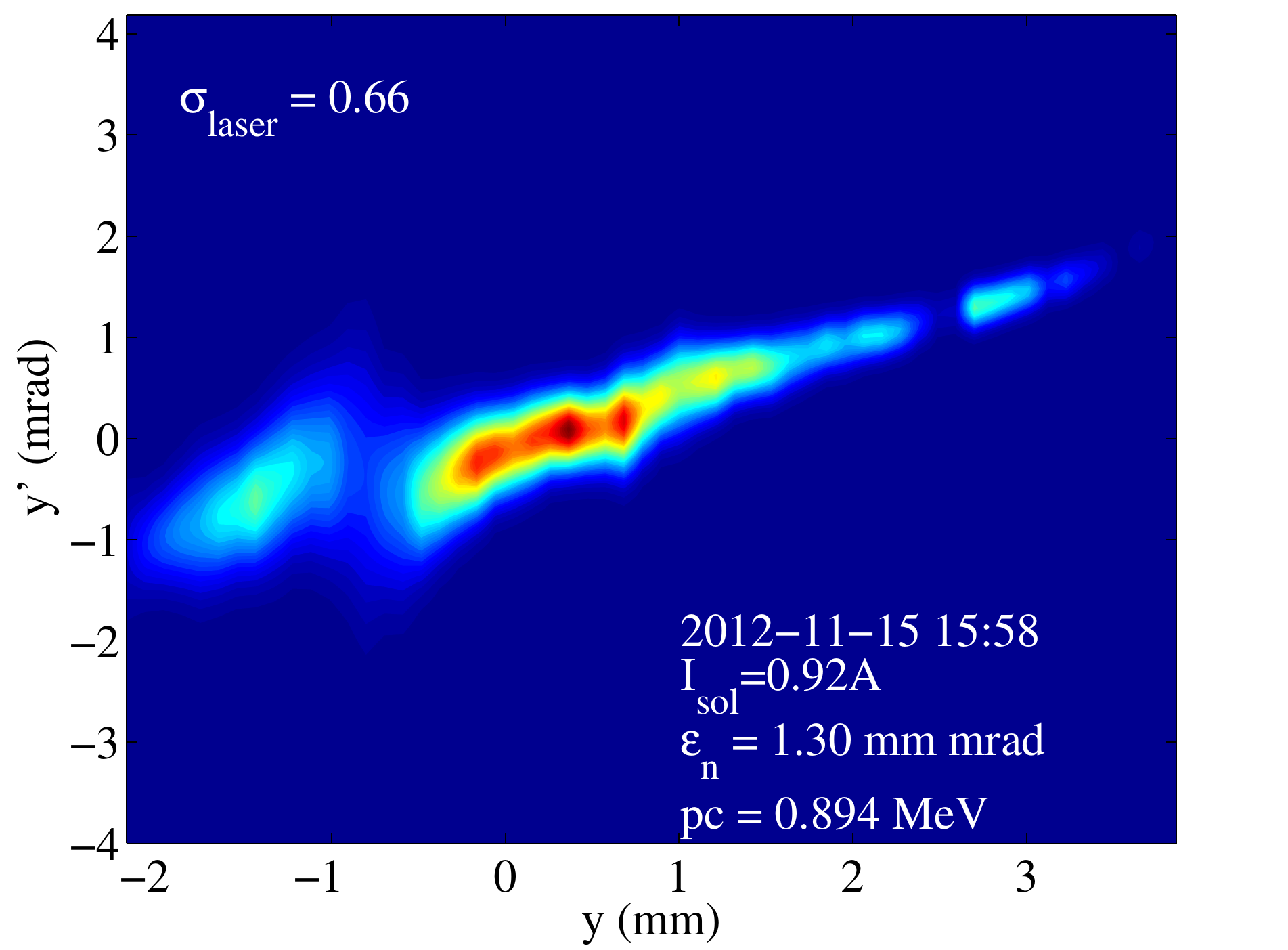}}
\caption{Reconstructions of vertical phase spaces, laser spot size increases in reading direction and is indicated in the figures.}
\label{fig:phase-spaces-laser}
\end{figure}

A linear dependency of the normalized emittance with respect to the rms laser spot size can be recognized, despite the scattering of the data. The corresponding phase spaces are displayed in figure \ref{fig:phase-spaces-laser}, where at large spot sizes a structured phase space was obtained. This hints at structured emission from hot spots on the cathode surface which may be covered with protrusions and droplets.

Uncertainties due to temporal deviations and other statistical errors are estimated to amount to less than 5\,\%. The finite thickness of the YAG screen together with a 45\,deg viewing angle may introduce an overestimation of the measured beam size of up to 30\,$\mu$m \cite{myBardayIBIC2012}. From a numerical estimate this will introduce a systematic error of about +3\,\% in the emittance. Space charge effects are negligible at the observed bunch charges below 1\,pC. Only these effects are taken into account by the error bars in figure \ref{fig:slit-mask-results}.

\begin{figure}[ht]
\vspace{-0.1cm}
\centering
\includegraphics[width=.45\textwidth]{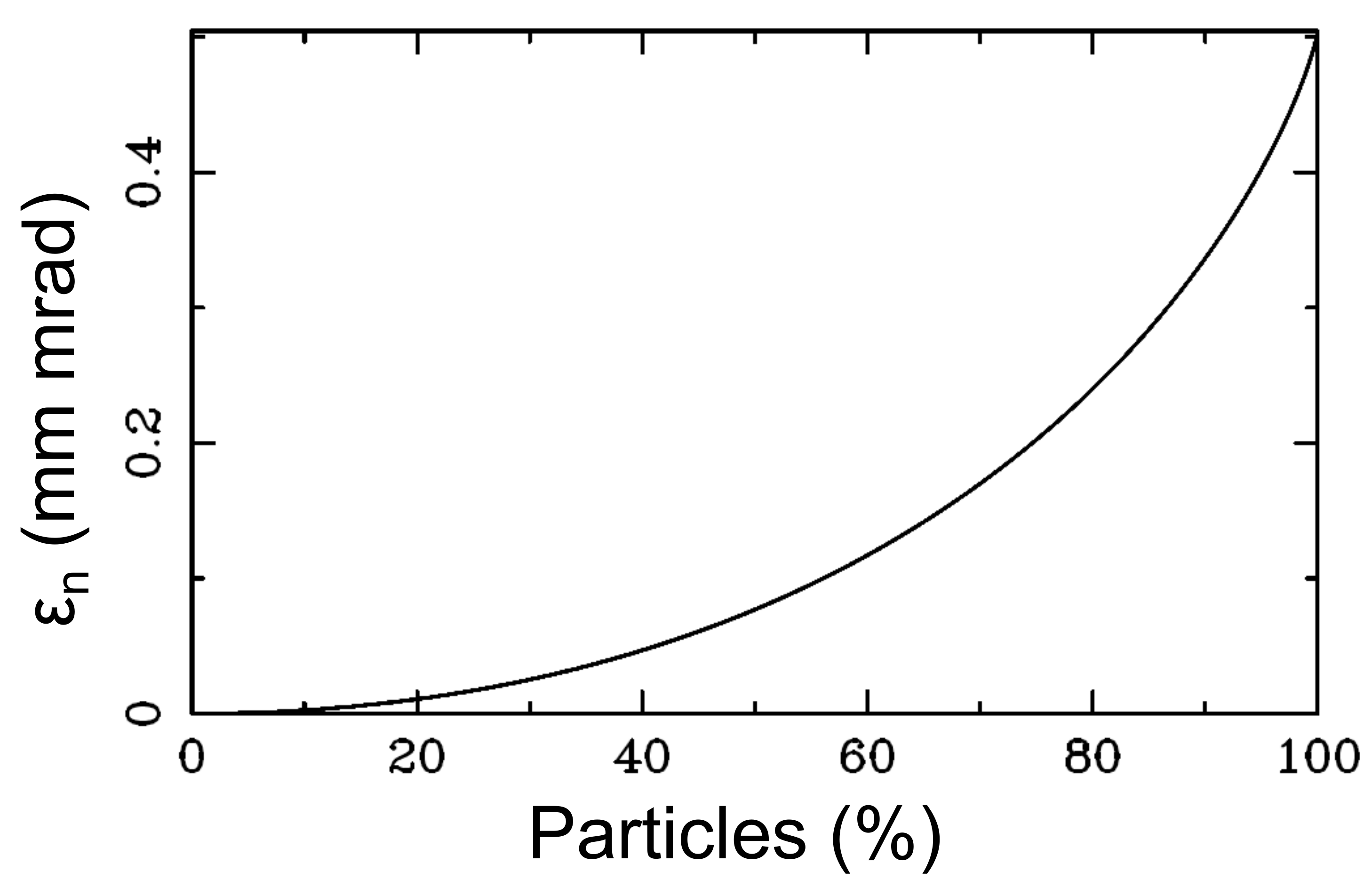}
\caption{Simulated charge cut curve of a round beam from the photoinjector.}
\label{fig:charge-cut}
\end{figure}

The accuracy of the evaluated emittance is largely defined by the dynamic range of the entire measurement, which differs between measurements because beamlets at the tail of the distribution might move off the screen due to geometrical constraints.
The range differs between 2.5 and 6.4\,dB, which implies that differences as large as
20\,\% in the number of considered particles occur.
For reference, a simulated charge cut curve in figure \ref{fig:charge-cut} shows the dependence of the emittance on the amount of particles imaged.

\begin{figure}[ht]
\centering
\mbox{\includegraphics[width=.5\textwidth]{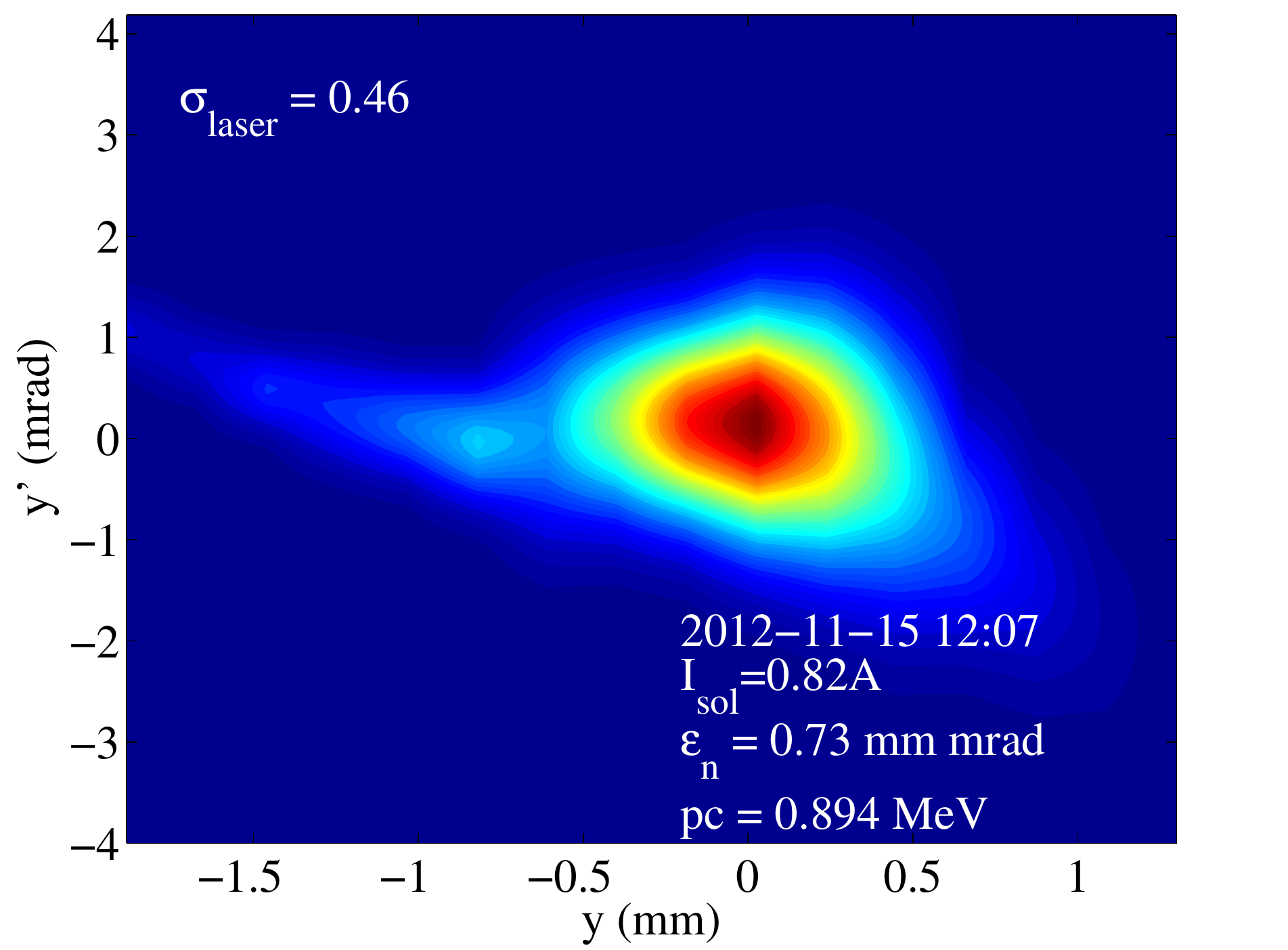}
\includegraphics[width=.5\textwidth]{PhaseSpace-920mA-FWHM}}
\includegraphics[width=.5\textwidth]{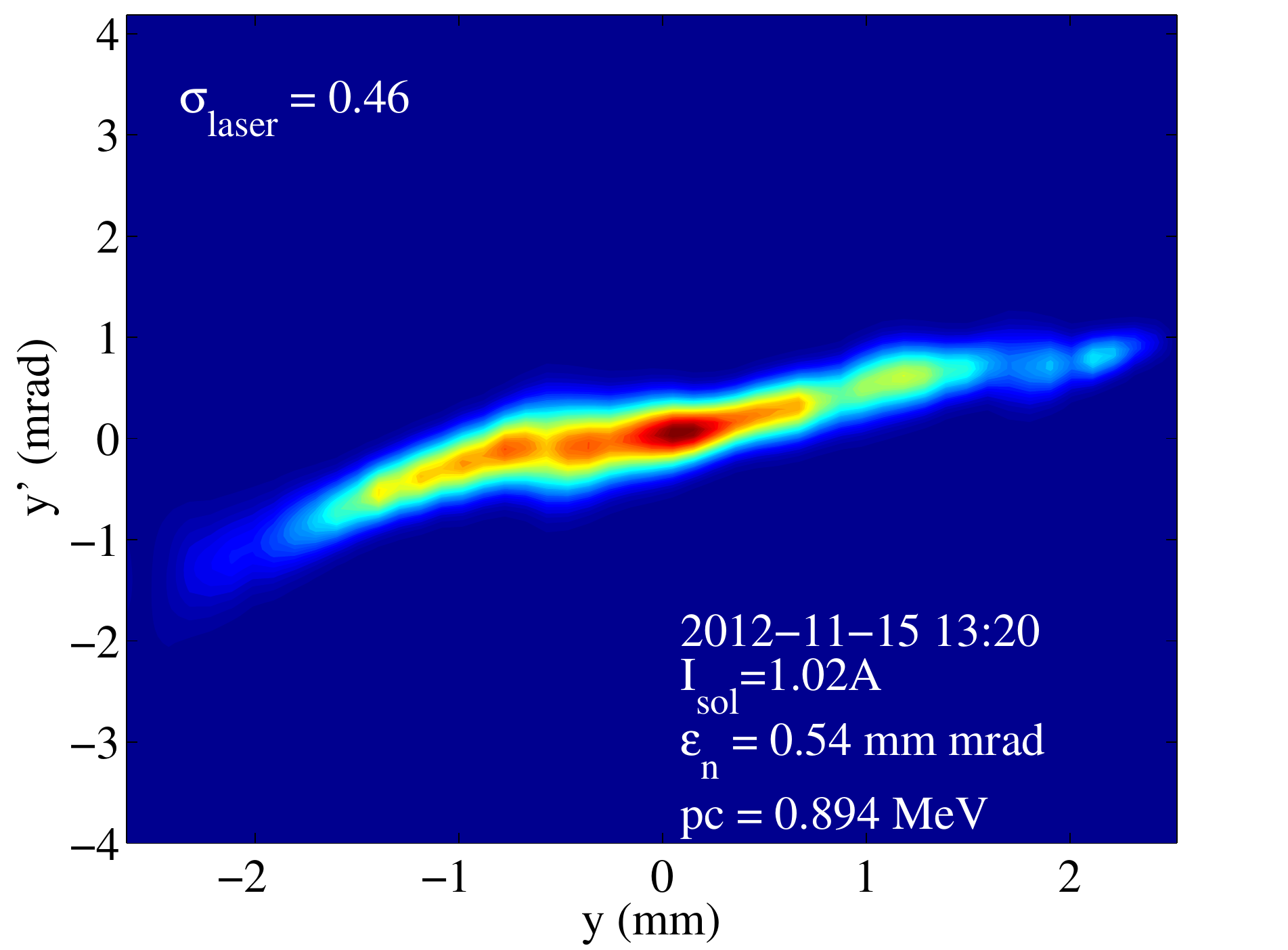}
\caption{Reconstructions of vertical phase spaces at increasing solenoid currents, as indicated in the figures.}
\label{fig:phase-spaces-sol}
\end{figure}

\clearpage
Three measurements with identical beam settings but increasing solenoid current were performed. The phase spaces are printed in figure \ref{fig:phase-spaces-sol}. The correlation between the vertical position and divergence is clearly increasing with the solenoid current as the focal point moves closer to the solenoid.

\section{Solenoid Scans}

\begin{figure}[!ht]
\centering
\includegraphics[width=.75\textwidth]{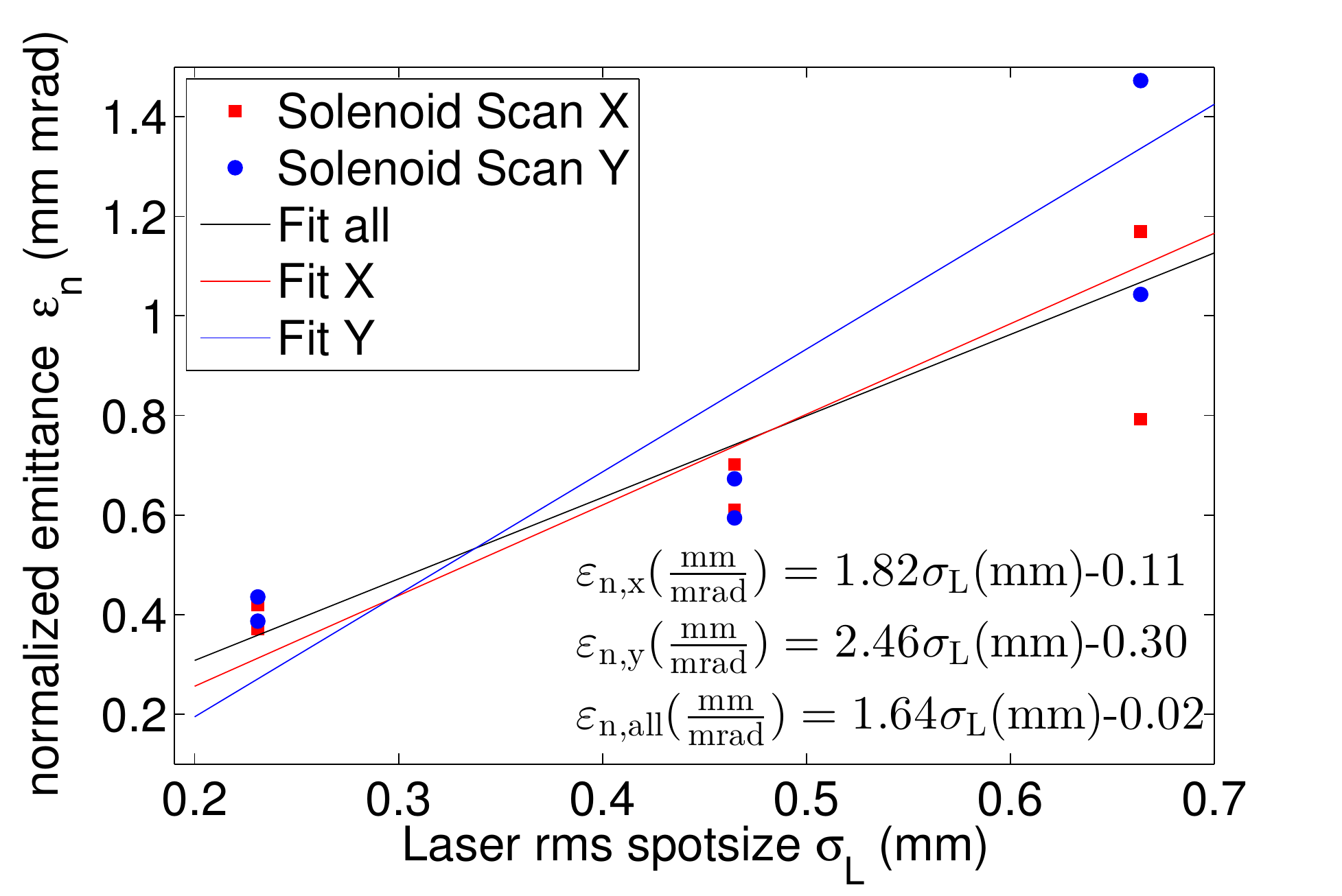}
\caption{Emittance measurements using the solenoid scan technique at different laser diameters. Bunch charges are not equal. The lower values are from measurements at 15\,deg launch phase, the higher ones at 25\,deg. Horizontal and vertical fits are for the 15\,deg measurement only.}
\label{fig:sol-scan-laser}
\end{figure}

Emittance measurements with different laser spot sizes were also carried out using the solenoid scan, see figure \ref{fig:sol-scan-laser}. The data was taken with a field amplitude of 10\,MV/m at launch phases of 15 and 25\,deg, which corresponds to beam energies of 0.94 and 0.90\,MeV, respectively. The beam was imaged on the front screen, at a distance of 1.121\,m to the center of the solenoid coil. As was observed with the slit mask measurements, the data shows a linear dependency, however, the values are slightly larger in this case.

\begin{figure}[ht]
\centering
\includegraphics[width=.75\textwidth]{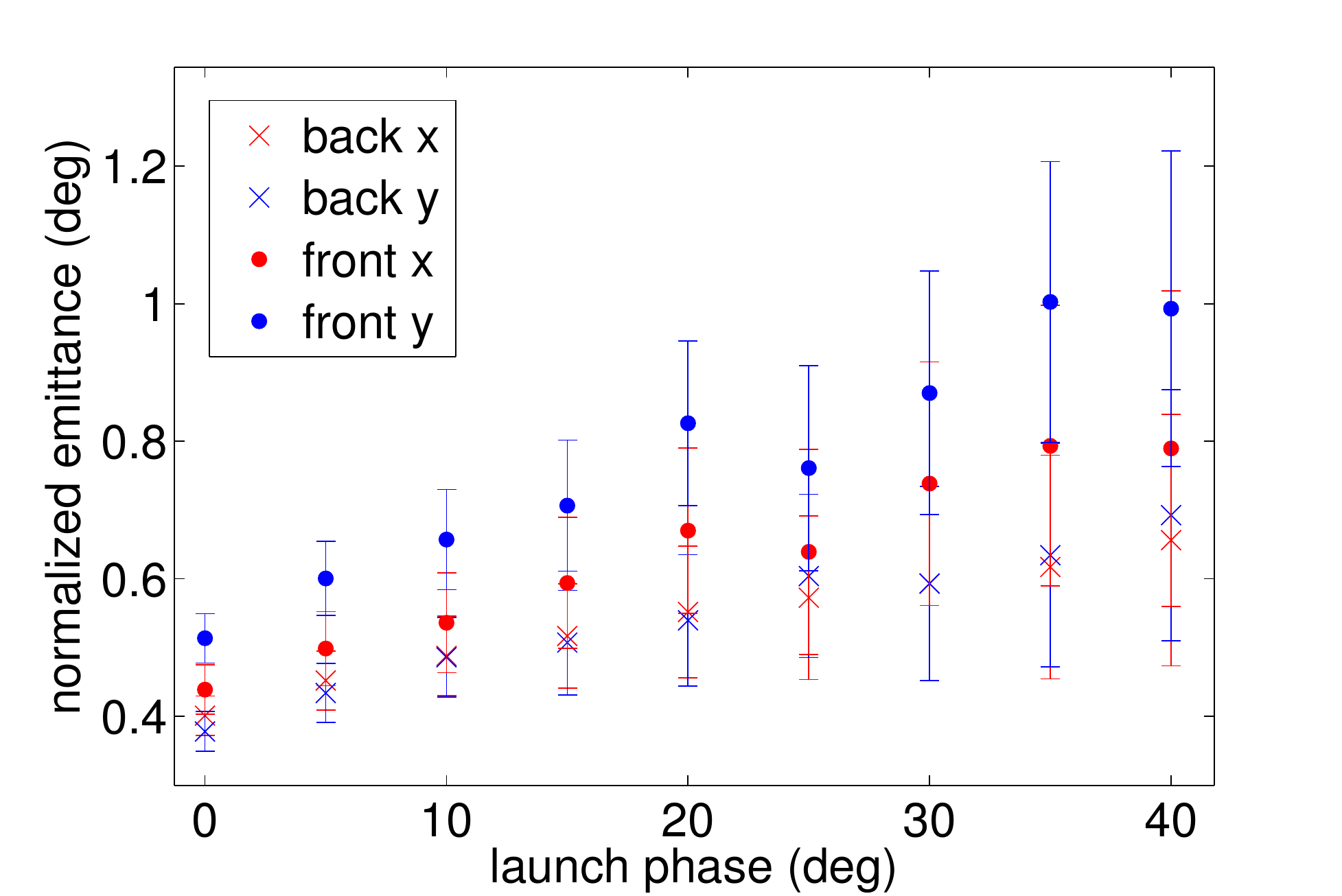}
\caption{Emittance measurements for different emission phases on the front and back screen. Error bars account for chromatic aberration and astigmatism.}
\label{fig:PhaseScan-emittances}
\end{figure}

\begin{figure}[ht]
\centering
\includegraphics[width=.75\textwidth]{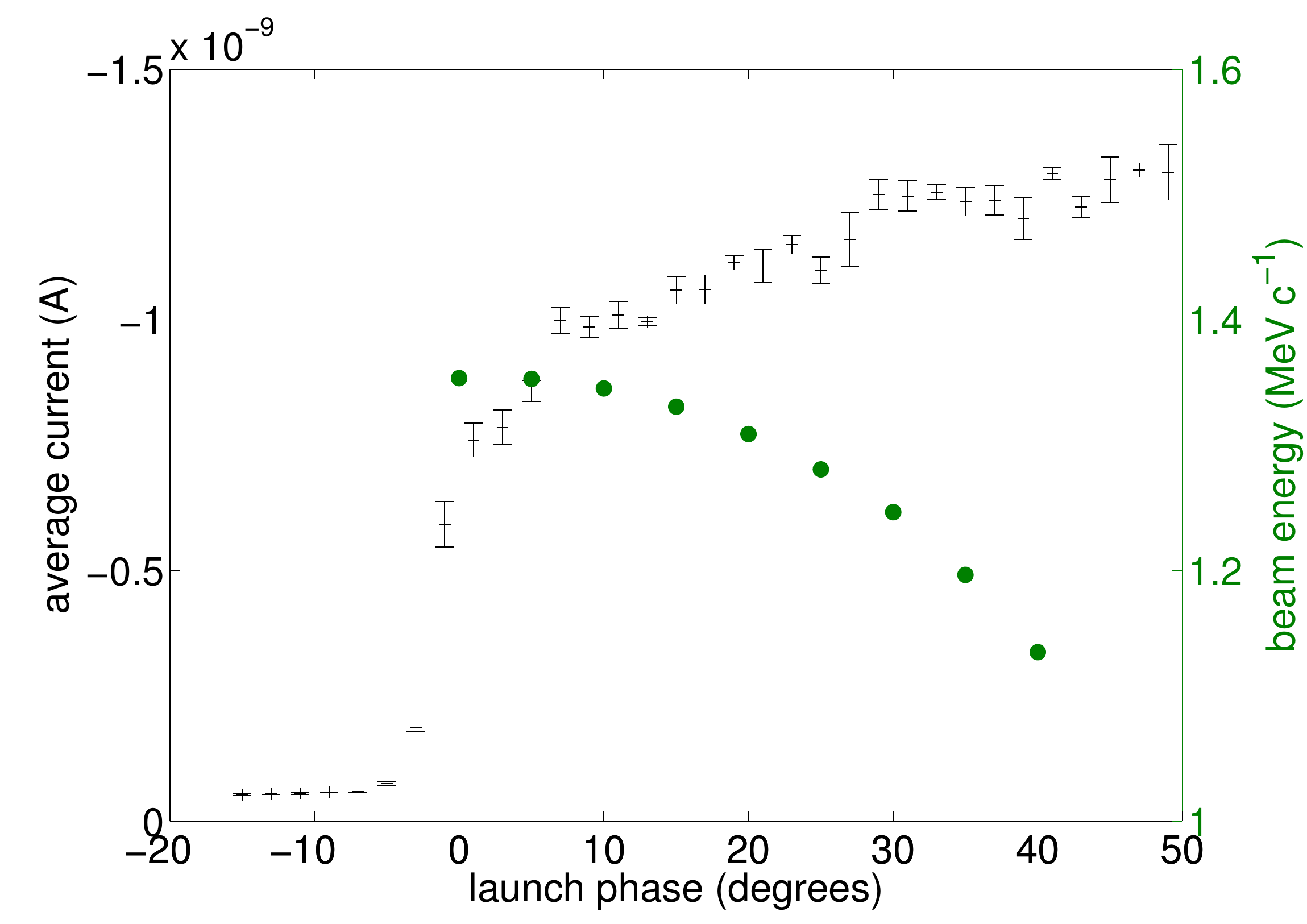}
\caption{Phase scan of the beam current and energy.}
\label{fig:PhaseScan-current-energies}
\end{figure}

Additionally, emittance measurements for different emission phases were conducted at a gradient of 12.5\,MV/m.
The resulting emittances are displayed in figure \ref{fig:PhaseScan-emittances}.
For reference, the corresponding average currents and beam energies can be found in figure \ref{fig:PhaseScan-current-energies}.
The emittance rises linearly with the launch phase up to 30\,deg and seems to flatten between 30 and 40\,deg.

Values measured using the front screen are clearly larger than the ones measured on the back screen.
Starting from 5\,\% at low launch phases the difference increases also linearly with the phase to up to 20\,\% at phases of 30 and 40\,deg.
Higher emittance values on the front screen, where the beam has to be focused stronger, suggest that strong focussing introduces aberrations that lead to an increase of the emittance.
The effect of chromatic and spherical aberrations of the solenoid was investigated with several simulations in \cite{myVolker2012}.
For similar beam parameters and the same solenoid it was shown that the spherical aberration coefficient $C_s$ increases from about 630~m$^{-3}$ to about 960~m$^{-3}$ when moving the focus from the back to the front screen.
The emittance increase due to spherical aberrations depends on higher-order moments of the particle distribution \cite{myVolker2012}:
\begin{equation}
\varepsilon^2 = \varepsilon^2_{(lin)} + 2 \, C_s\left(\langle r^3_i r_i^\prime \rangle \langle r_i^2 \rangle - 3 \langle r_i r_i^\prime \rangle\langle r_i^2 \rangle^2 \right) + 6 \, C_s \, \langle r_i^2 \rangle^4
\end{equation}
which are not known, however a rough estimate with values from simulations and the slit mask measurements yields an emittance increase that is several orders of magnitude lower than the measured difference.

\begin{figure}[ht]
\centering
\includegraphics[width=.75\textwidth]{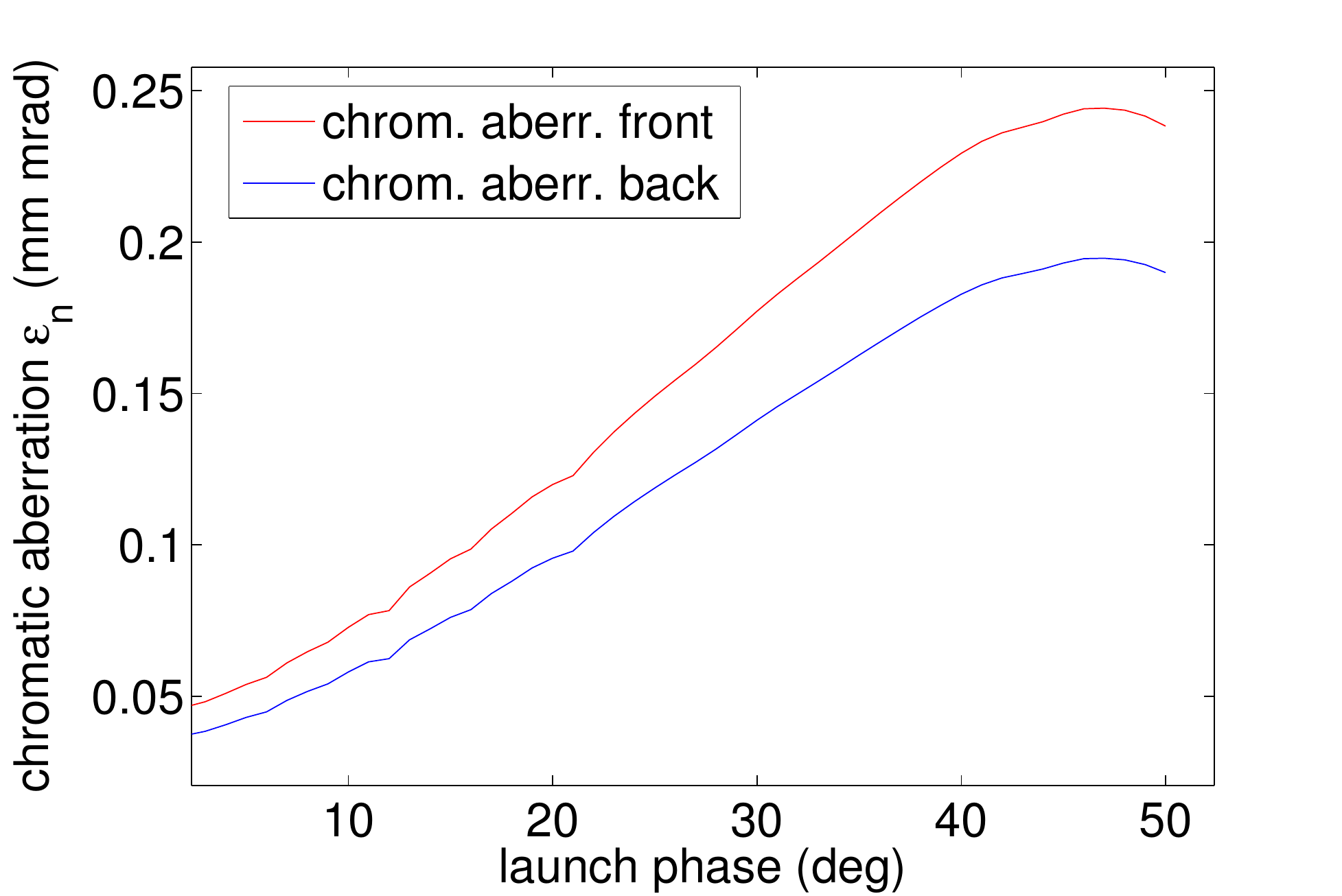}
\caption{Influence of the chromatic aberration of the solenoid on the beam emittance for a beam at 12.5\,MV/m accelerating gradient. Front and back refers to solenoid settings where the beam is focused on the front and back screen, respectively.}
\label{fig:chromaticity}
\end{figure}

The chromatic aberration follows \cite{Dowell}
\begin{equation}
\varepsilon_{n,chrom} = \sigma^2_{x,sol} \, \frac{\sigma_p}{mc} \, K |\sin KL + KL \cos KL|
\end{equation}
where $\sigma_{x,sol}$ is the rms beam width at the solenoid, $\sigma_p$ is the rms momentum spread, $K=\frac{eB_0}{2p}$ and $L$ is the solenoid's effective length. Because no vertical slit was installed in front of the dipole, a precise measurement of the momentum spread was not possible. The influence of the chromatic aberration was estimated using beam size and momentum spread values from Astra simulations. As can be seen in figure \ref{fig:chromaticity}, the influence is much stronger when focusing the beam on the front screen. Overall, the chromatic aberration may be responsible for the increase of the measured emittance with the launch phase as it also increases nearly linearly below 30\,deg and flattens above 30\,deg. Additionally, it explains the discrepancy between the measurements on the front and back screen, because focusing on the front screen introduces higher aberrations due to the stronger solenoid field.

A third source of uncertainty is caused by the astigmatism of the solenoid. It causes two separated foci for the two transverse directions. When the beam is focused in $x$ direction, it may still have twice its minimal size in $y$ direction and vice versa. Because the beam is rotated by the larmor angle $\Theta_L$ with respect to the lab frame, this may result in a vast overestimation of the beam diameter on the screen and, subsequently, of the reported emittance. This effect is especially prominent in the measurements on the front screen. It is assumed that the difference between the horizontal and vertical emittance in the measurements on the front screen is due to the astigmatism. The lower uncertainty level was estimated to
\begin{equation}
\Delta \varepsilon = C \, \sigma_y\bigg|_{\sigma_x = min} \, \sin(\Theta_L)
\end{equation}
where $\sigma_y$ is the beam diameter in $y$ direction at the location of the beam waist in $x$ direction. C is a constant and was set to 0.25 for the present results. It is somewhat arbitrary because the influence of the larmor rotation on the projected intensity is not known for the specific distribution.
Chromatic aberration and astigmatism effects are accounted for in the error bars of figure \ref{fig:PhaseScan-emittances}.

\chapter{Conclusion and Future Outlook} \label{chap:Conclusion}

\section{Performance of the Gun}

Gun 0.2 demonstrated production of a low current electron beam at energies between 1 and 2.5\,MeV. The normalized emittance was measured to be 1.9\,mm\,mrad per mm rms laser spot size by the slit mask technique and 1.8\,mm\,mrad\,/\,mm rms (horizontal) and 2.5 \,mm\,mrad\,/\,mm rms (vertical) by the solenoid scanning technique. Overall, the performance has much improved over the previous setup, where the normalized emittance was between 5.2 and 5.7\,mm\,mrad / mm rms. This is probably due to better control of the lead deposition process on the cathode plug. Visually, the beam was less structured than on images from the previous setup.
The QE for cavity 0.1 is higher than for cavity 0.2 by a factor of 10 because laser cleaning with an excimer laser of the Pb cathode film was only performed with cavity 0.1 \cite{myBardayIPAC2013}.
Lower available average drive laser power further reduced the average current generated from the cathode by a factor of 3.
Generally, the plug-gun concept offers advantages over the direct coating of the inner back wall. Lower dark current and lower emittance offer high performance characteristics for the hybrid Nb/Pb gun.

\section{Consequences for Future Measurement Setups}

There are several features of the collected data that require further investigation in order to identify the causes.
Currently, the data does not allow to quantify the influence of cavity and emission dynamics and the influence of dynamic range differences or distinguish between them. Thus, it was not possible to clearly attribute the discrepancy between the results of slit mask and solenoid scan measurements to one of them. The effect of solenoid aberrations on both measurements could be mitigated by weak focusing, which implies that the back screen should be used for solenoid scans and a larger screen is required to image all beamlets when a large beam diameter is scanned over the slit aperture for phase space characterization.
The emittance increase with the launch phase can  be attributed to the chromatic aberration of the solenoid.

Regarding the measurements themselves, future ones should be conducted with greater care for comparability. Bunch charges should be kept equal when changing the laser diameter. Emission phases and the field gradient in the cavity should be equal to allow comparison of different measurement techniques.
The back screen is currently in a position where a sharp image of the cathode's emission surface is created when the beam is focused on the slit mask.
A vertical slit in front of the dipole is required to make meaningful measurements of the momentum spread.

A setup to directly measure the phase space in short time (about 1~min) using a double slit design has been developed at Cornell \cite{Bazarov2008}. The entirely electric measurement delivers a higher and more reliable dynamic range. Additionally, faster emittance measurement would allow to better correlate gun parameters and beam emittance.
For further investigations a momentatron device is planned that directly images the intrinsic transverse momentum distribution of the cathode after only an acceleration towards a grid and a short drift space \cite{myVecchione2011}.

\begin{appendix}
	\chapter{Contribution to IPAC 2013}
	\includepdf[pages={-}]{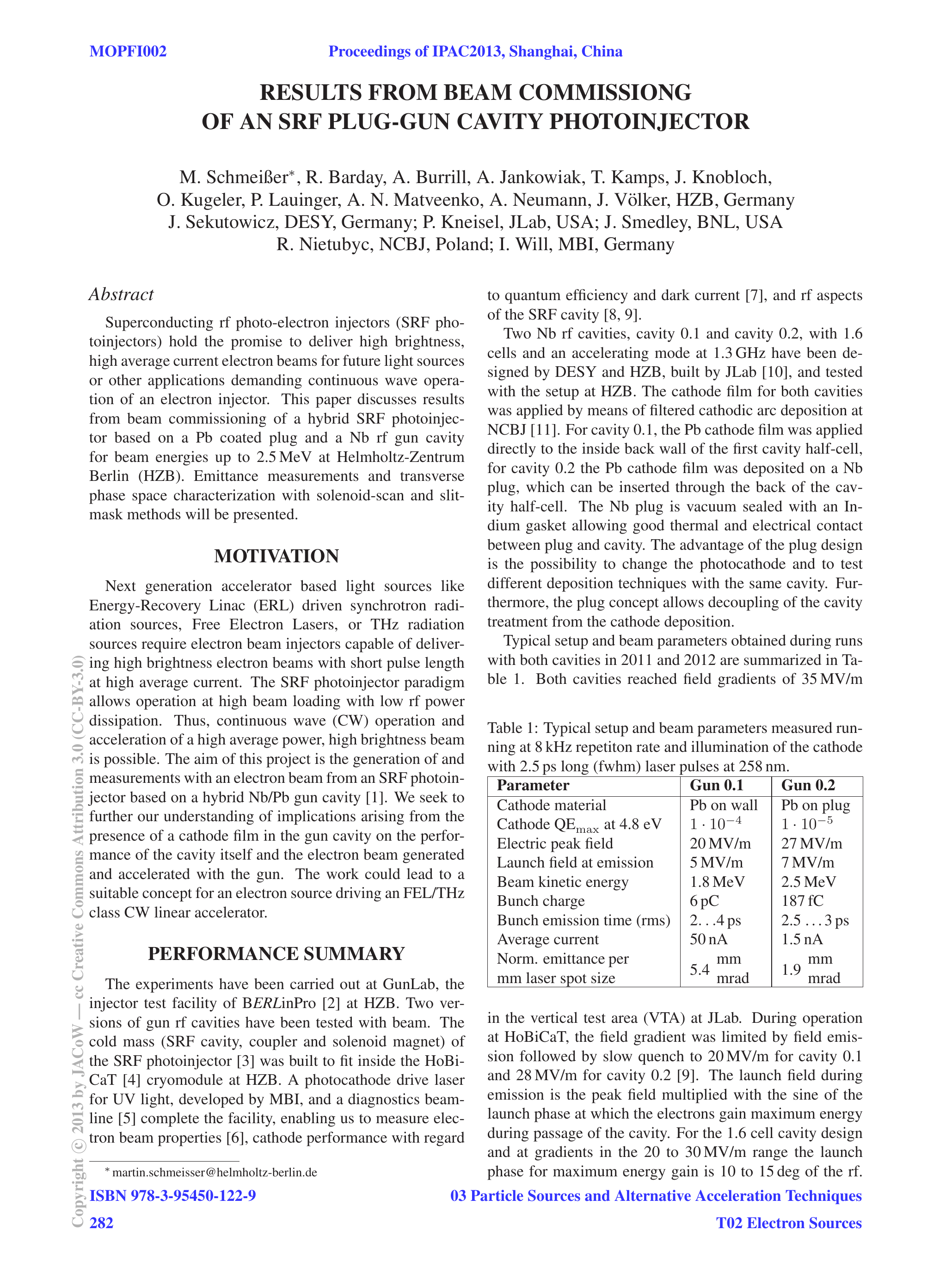}
	\includepdf[pages={-}]{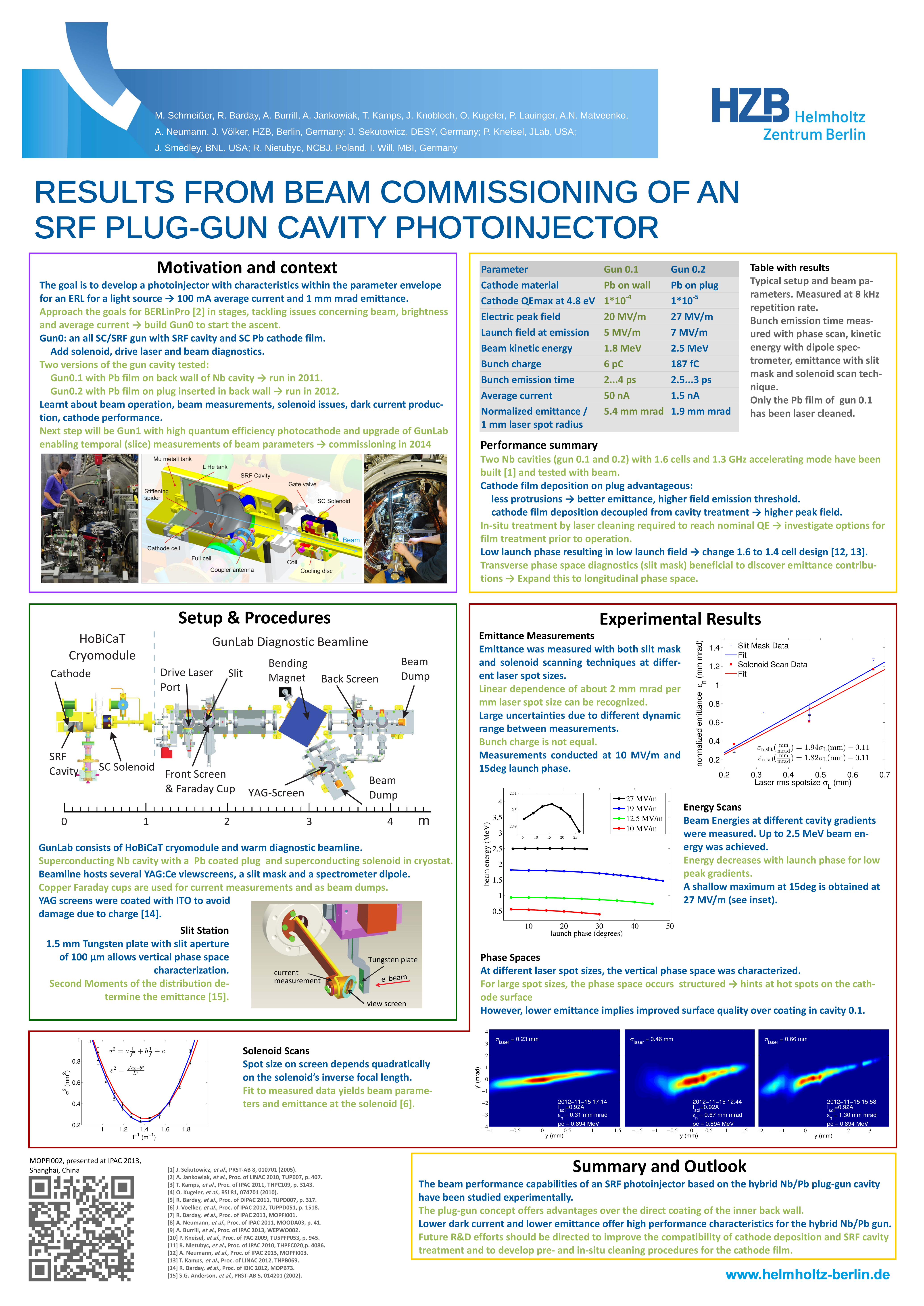}
\end{appendix}

\bibliographystyle{unsrtnat}
\bibliography{library,mylibrary}

\end{document}

%% file: metadata.tex

\newcommand{\dcsubject}{Forschungsbeleg}
\newcommand{\dctitle}{Emittance Measurements of a Superconducting High Frequency Electron Gun}
\newcommand{\dcsubtitle}{}

\newcommand{\dcauthorlastname}{Schmei{\ss}er}
\newcommand{\dcauthorfirstname}{Martin}
\newcommand{\dcauthormiddlenames}{Anton Helmut}
\newcommand{\dcauthoremail}{martin.schmeisser@physik.hu-berlin.de} 
\newcommand{\dcdate}{\today}

\newcommand{\dcplace}{Berlin} 
\newcommand{\dcuni}{Humboldt Universit{\"a}t zu Berlin}
\newcommand{\dcunidepart}{Mathematisch-Naturwissenschaftliche Fakult{\"a}t I, Institut für Physik}
\newcommand{\dcunilogo}{husiegel_bw_op}
\newcommand{\dcinstitute}{Helmholtz-Zentrum Berlin f{\"u}r Materialien und Energie GmbH}
\newcommand{\dcinstdepart}{Institut f{\"u}r Beschleunigerphysik}
\newcommand{\dcinstlogo}{hzb_logo_cmyk}
\newcommand{\dcprof}{} 

\newcommand{\dcpruefer}{}
\newcommand{\dcsecpruefer}{}
\newcommand{\dcadvisor}{}

\newcommand{\dckeywords}{}

\hypersetup{%
	pdftitle	= {\dctitle}, %
	pdfsubject	= {\dcsubject, \dcdate}, %
	pdfauthor	= {\dcauthorfirstname~\dcauthorlastname, \dcauthoremail}, %
	pdfkeywords	= {\dckeywords}, %
	pdfcreator	= {pdfTeX with Hyperref and Thumbpdf}, %
	pdfproducer	= {LaTeX, hyperref, thumbpdf}, %
}

%% file: title.tex

\titlehead{

\vspace{1cm}

\parbox{6cm}{\centering \includegraphics[width=4cm]{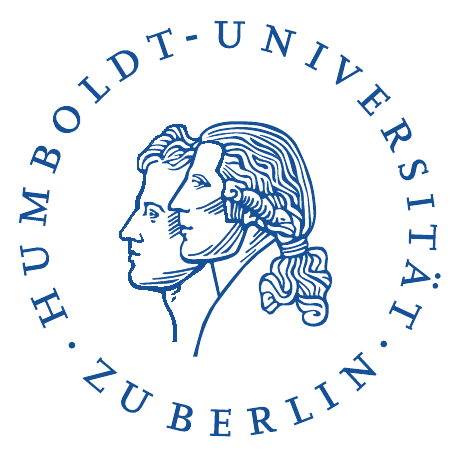} } 
\hfill
\parbox{6cm}{\centering \includegraphics[width=6cm]{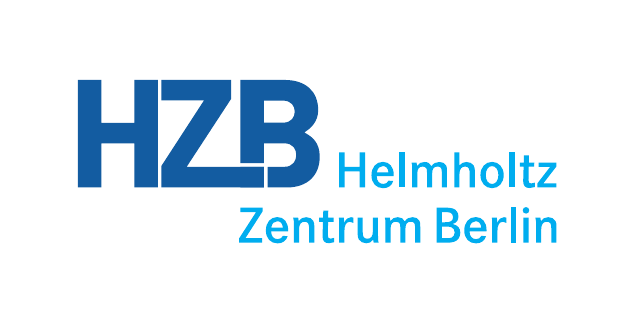} }
\hfill

\vspace{2cm}
}

\subject{\bf\Huge\dcsubject}


\title{\Large
	\dctitle
	\dcsubtitle
}

\author{\dcauthorfirstname~\dcauthorlastname}
	
\selectlanguage{ngerman}
\date{\dcplace, \dcdate
\selectlanguage{english}
}


\lowertitleback{
\small

\textbf{\dcauthorlastname, \dcauthorfirstname~\dcauthormiddlenames}

\dctitle~-~\dcsubtitle

\dcsubject

\dcuni\\
\dcunidepart

\dcinstitute\\
\dcinstdepart

\selectlanguage{ngerman}
\monatwort{\the\month}~\the\year
\selectlanguage{english}
}

\maketitle

\cleardoubleemptypage

\selectlanguage{english}
\pagenumbering{roman}
\tableofcontents

\cleardoublepage
\pagenumbering{arabic}
\setcounter{page}{1}